\documentstyle[aps,prb,twocolumn,epsfig]{revtex}
\begin{document} 
\draft

\makeatletter
\global\@specialpagefalse
\def\@oddhead{\hfill submitted to Physical Review Letters}
\let\@evenhead\@oddhead
\def\@oddfoot{}
\let\@evenfoot\@oddfoot
\makeatother

\wideabs{

\title{ Anomalous quantum confined Stark effects in stacked InAs/GaAs self-assembled quantum dots } 
\author{ Weidong Sheng and Jean-Pierre Leburton } 
\address{ Beckman Institute for Advanced Science and Technology and Department of Electrical
and Computer Engineering,\\University of Illinois at Urbana-Champaign, Urbana, Illinois 61801 }
\vspace{-8.25mm}
\rightline{submitted to Physical Review Letters}
\vspace{+8.25mm}
\maketitle

\begin{abstract} 
Vertically stacked and coupled InAs/GaAs self-assembled quantum dots (SADs) are predicted to
exhibit a strong non-parabolic dependence of the interband transition energy on the electric
field, which is not encountered in single SAD structures nor in other types of quantum
structures. Our study based on an eight-band strain-dependent ${\bf k}\cdot{\bf p}$
Hamiltonian indicates that this anomalous quantum confined Stark effect is caused by the
three-dimensional strain field distribution which influences drastically the hole states in
the stacked SAD structures.
\end{abstract}

}\narrowtext

Zero-dimensional semiconductor structure, like InAs /GaAs self-assembled quantum dots (SADs)
\cite{bim} have attracted considerable attention because of the new physics \cite{ebi,sti,keg}
of a few electron systems and potential applications in optoelectronics \cite{dep}. Recent
experiment on Stark effect spectroscopy in SADs \cite{fry} has demonstrated the existence of
an inverted electron-hole alignment due to the presence of gallium diffusion in InAs SADs, and
established a relation between the Stark shift and the vertical electron-hole separation.

The theoretical interpretation of these experimental results is based on the assumption that
the applied electric field can be treated by the second-order perturbation theory, which
results in a quadratical dependence of the transition energy on the applied electric field
\cite{bar}, \begin{equation} E(F) = E(0) + p F + \beta F^2 \end{equation} where $p$ is the
built-in dipole moment and $\beta$ measures the polarization of the electron and hole states,
{\it i.e.}, the quantum confined Stark effect. While this relation is well satisfied in many
quantum systems including single SADs \cite{bar}, and quantum well structures \cite{bas,mat},
we show in this work that it is not valid for vertically coupled SAD structures \cite{eis}
where the quantum confined Stark effect deviates significantly from its quadratic dependence
on the electric field. The reason for this anomalous quantum confined Stark effect is due to
the three-dimensional (3D) strain field distribution in the dots and in the coupling region,
which controls the localization of hole states in the respective SADs, and their sensitivity
to external field. The existence of this effect is important for basic condensed matter
physics because it can not be inferred from a simple superposition of the electronic properties
of single SADs. It is also promising for applications in optoelectronics because interband
transition energies can be significantly modulated by electric fields in quantum dot lasers
and other photonic devices.

The insets of Fig.~1 show schematically a single SAD structure and a system of two vertically
coupled SADs that are truncated pyramids separated by a GaAs barrier of $1.8$~nm, with
identical base $17.4\times 17.4~nm^2$ and individual height $3.6$~nm. A positive electric
field $F$ is directed from the top to the bottom of the structures. Fig.~1 shows the
calculated ground state transition energies for the single dot and for the stacked structure,
as functions of electric fields. The electron and hole states of the system are obtained from
the Schr\"{o}dinger equation in the framework of the envelop function formalism \cite{bah},

\begin{equation} 
({\bf H}_{k\cdot p} + |e|Fz) \phi = E \phi.
\label{eq:H} 
\end{equation} 
Here $\phi=(\phi_1, \phi_2, \ldots, \phi_8)$ is the envelop eigenvector and ${\bf H}_{k\cdot
p}$ is the $k\cdot p$ eight-band Hamiltonian which includes the effect of strain and
piezoelectricity \cite{fon,pry}. The Hamiltonian is discretized on a three-dimensional grid as
a large sparse matrix which is solved by Lanczos algorithm. This approach has been shown to be
reliable, especially in the investigation of the inverted electron-hole alignment in SAD
structures \cite{swd}.

\begin{figure}[h]
\centerline{\psfig{file=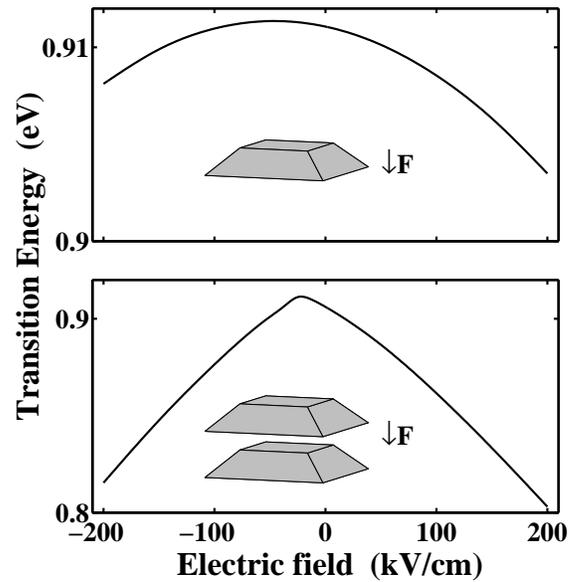,width=3.25in}}
\caption{ \small Ground state transition energies as functions of vertical electric fields for
single dot (top) and stacked double dots (bottom). Insets show the dot systems schematically.
}
\end{figure}

\begin{figure}[h]
\centerline{\psfig{file=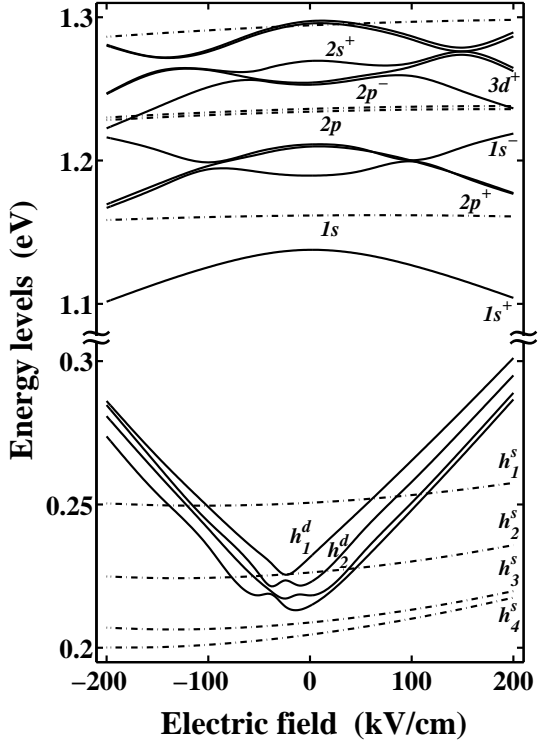,width=3.25in}}
\caption{ \small Energy levels of stacked double dots, as a function of the electric field.
The dash-dotted lines are for the single dot of the same size. }
\end{figure}

The single dot exhibits a nearly perfect quadratical dependence on the electric field which is
referred to the conventional quantum confined Stark effect and has been observed in many other
types of quantum structures \cite{bar,bas,mat}. The maximum transition energy occurs at
$F\approx -50$~kV/cm because this homogeneous InAs SAD has a positive electron-hole alignment,
{\it i.e.}, the hole is at the bottom of the dot for zero electric field. The stacked
structure shows piecewisely quasi-linear dependence, with a turning point at a smaller
negative electric field $F\approx -22$~kV/cm. The lower panel of Fig.~1 shows that this
discrepancy from the conventional quantum confined Stark effect occurs over the whole range of
electric field, {\it i.e.}, for $200$~kV/cm~$\leq F\leq $~$200$~kV/cm. In addition, the
stacked structure exhibits a Stark shift of one order of magnitude stronger than the single
dot. We also notice that the maximum transition energy for both systems are roughly identical,
because in the stacked structure, despite the downshift of the electron states due to tunnel
coupling, the hole states are characterized by an up-shift in energy due to the strain
distribution (see Fig.~2). Similar simulations have been performed for stacked structures with
different sizes, different thicknesses of the coupling region, and with or without wetting
layers, where the same kind of anomalous quantum confined Stark effect has been found
\cite{jpl}.

Fig.~2 shows the energy levels of the stacked structure, for both electrons and holes, as a
function of electric fields. All the energies are given in reference to the top of the valence
bands of GaAs. For the sake of showing the fine structure of the hole spectra near zero
electric field, we use an energy scale twice as small as for electrons. For comparison, the
energy levels of the single dot, are shown in dash-dotted lines. It is clearly seen that, due
to the coupling between the stacked structures, the double dot system exhibits much richer
structure in its energy spectrum.

In the conduction band, the energy spectrum of the single dot shows the ground {\it 1s} state,
the two nearly degenerate excited {\it 2p} states, and the {\it 3d}-like state with weak
sensitivity to the electric field. The coupled dot system results in bonding and antibonding
states originating from these states, that in the absence of electric field, are identified as
$1s^+, 1s^-, 2p^+, 2p^-, 2s^+, 3d^+, \cdots$ where $+(-)$ denotes bonding (antibonding)
states. Except for the ground state, all the excited states are seen to have crossings or
anticrossings with other states, which reorders the states at high fields. In the valence
bands of the single dot, the four hole levels ($h_1^s$, $h_2^s$, $h_3^s$ and $h_4^s$) are seen
to show quadratical dependence on the electric field.

The most dramatic feature in the valence bands of the stacked structure is the quasi-linear
dependence of the hole levels on the electric field for intermediate and strong field
intensities, especially the ground hole state $h_1^d$. In addition, we could not find any
bonding or antibonding hole states for these structure as the ground hole state $h_1^d$ and
first excited state $h_2^d$ are localized in the bottom dot and the top one, respectively.
However, at small electric fields, the hole levels show small fluctuations due to mutual
anticrossings. It is also noticed that the magnitude of the Stark shift for holes in the
stacked structure is significantly larger than in the single dot, which is responsible for the
large shift in the transition energy shown in Fig.~1.

The main reason for the hole states to behave so different in the stacked structure than in
the single dot is traced in the 3D strain field distribution. Fig.~3(a) shows the profiles and
the corresponding contour plots (insets) for the hydrostatic (H) and the biaxial (B)
components of the strain field, which are defined as

\begin{eqnarray}
&H = \epsilon_{xx} + \epsilon_{yy} + \epsilon_{zz}& \nonumber \\ 
&B^2 = (\epsilon_{xx}-\epsilon_{yy})^2 + (\epsilon_{yy} - \epsilon_{zz})^2 + (\epsilon_{zz} - \epsilon_{xx})^2&
\end{eqnarray}
Fig.~3(b) shows the diagrams of the conduction and valence band edges which are given by
\cite{bah2}

\begin{eqnarray}
&\Delta U_c = \Delta E_c + a_c \cdot H& \nonumber \\
&\Delta U_v = \Delta E_v - a_v \cdot H - b \cdot B / \sqrt{2}& 
\end{eqnarray}
where $\Delta E_c$ ($\Delta E_v$) is the conduction (valence) band offset, $a_c$ ($a_v$) is
the conduction (valence) band hydrostatic deformation potential parameter, $b$ is the
valence band shear deformation potential parameter. Hence, while electrons are only
sensitive to hydrostatic strain, holes are also and mostly affected by biaxial strain. 

\begin{figure}[h]
\centerline{\psfig{file=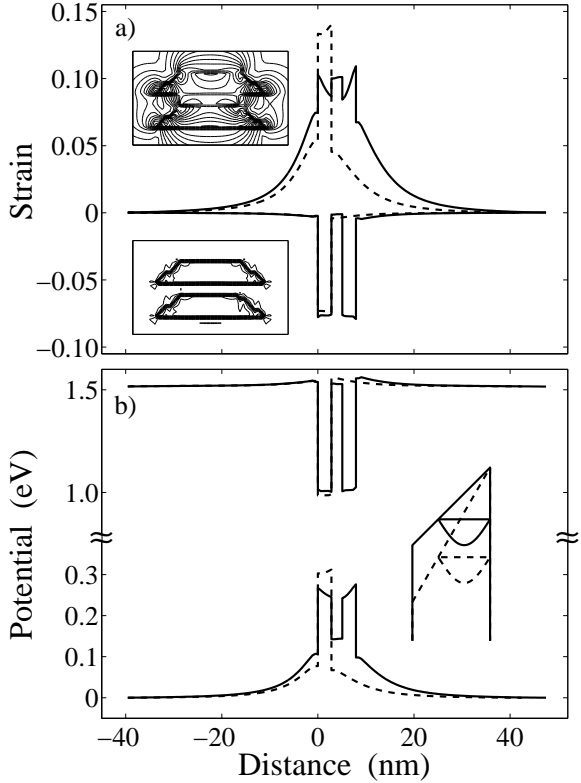,width=3.25in}}
\caption{ \small (a) Biaxial (upper curve) and hydrostatic (lower curve) components of the
strain field in the single (dash lines) and stacked (solid lines) SAD structures through the
center of the dots along the growth direction (z). The zero on the horizontal axis is fixed at
the top of the upper dot. (b) Band diagrams for the single (dash lines) and stacked (solid
lines) SAD structures. Inset: Schematic view of the ground hole state confined by a triangular
potential in the bottom dot at zero electric field (solid lines) and a positive electric field
(dash lines). }
\end{figure}

In the stacked structure, the hydrostatic negative strain field resides entirely inside the
dots, and is a little stronger than in the single dot, while it almost vanishes in the
coupling region (lower curve in Fig.~3a). Therefore, the coupling region is seen by electrons
as a conventional tunneling barrier, which results in bonding and antibonding states as shown
in Fig.~2. The biaxial positive strain field is however seen very differently in the stacked
structure from that in the single dot (upper curve in Fig.~3a). First, it is smaller than in
the single dot, leading to hole levels with higher energies in the stacked system (see
Fig.~1). Second, unlike the hydrostatic strain, the biaxial strain retains a substantial value
in the coupling region in the stacked system (top inset of Fig.~3a), which noticeably reduces
the barrier height in the valence bands. Third, the biaxial strain profile in the stacked
structure is inverted in the upper dot compared to the lower one although symmetric with
respect to a median plane between the dots.

\begin{figure}[h]
\centerline{\psfig{file=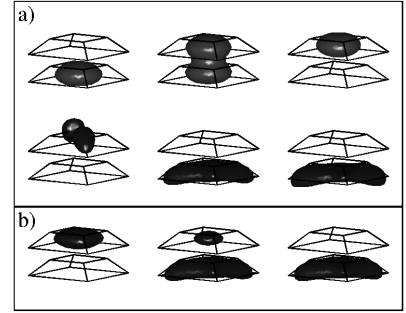,width=3.25in}}
\caption{ \small Probability density isosurfaces of ground states of electron (top panel) and
hole (two bottom panels) at different electric fields. From left to right, the electric field
is (a) $F=-200$, $0$, and $+200$~kV/cm; (b) $F=-21.8$, $-21.6$, and $-21.4$~kV/cm,
respectively. }
\end{figure}

The most important point is that the biaxial strain defines two triangular confining
potentials in each dot of the stacked structure, while in the single dot the valence band edge
profile exhibits a much smoother slope (see Fig.~3b). It is seen that the strain-induced
triangular potential is inverted with respect to the base in the two dots: In the bottom dot,
the valence band edge is higher at the base than at the top, while it is opposite in the top
dot. Therefore, the favorable combination of the lower strain-induced potential and the wider
base result in a weaker hole confinement in the bottom dot than in the top dot, thereby
localizing the hole ground state in the lower dot. The inset of Fig.~3(b) schematically shows
the potential in the lower dot of the stacked structure together with the ground hole state at
different electric fields. At zero electric field, the strain-induced built-in potential
variation in the dot is $34$~meV and the ground hole state is localized approximately one
fourth of the lower dot height from the bottom base. 

This situation is illustrated in the middle panel of Fig.~4(a) where we plot the probability
density isosurfaces of ground electron and hole states at different electric fields. In
contrast, it is seen that the ground electron state is the bonding $1s^+$ state extending
quasi-equally in both dots and in the coupling region. Under strong positive electric field,
the ground electron state is seen in the right panel of Fig.~4(a) to become a $1s$-like state
entirely localized in the upper dot. However, the ground hole state is found to behave very
differently. Because the hole state is localized inside this triangular potential, as
schematically shown in the inset of Fig.~3(b), its energy level changes approximately by the
same amount as external variation in the dot, {\it i.e.}, the potential drop arising from the
external electric field \cite{ft1}. For example, the variation of ground hole state energy
between $F=200$~kV/cm and zero field is about $70$~meV, which is about the same value as the
potential drop over the bottom dot caused by electric field $F=200$~kV/cm. Therefore it is
seen in Fig.~2 that the energy of the hole state has an almost linear dependence on the
electric field. This also explains the quasi-linear dependence of the ground state transition
energy as shown in Fig.~1.

In Fig.~3(b), the ratio between the tunneling barrier and band offset in the valence bands is
much smaller than that in the conduction bands. Consequently, it is much easier for holes to
tunnel through the barrier at negative electric fields than electrons. Fig.~4(b) illustrates
the transition of the ground hole state from the bottom dot to the top one occuring at a small
negative electric field $F=-21.6$~kV/cm. As the biaxial strain field has different
distribution in the respective dots, the hole states exhibit different probability density
profiles when localizing in different dots. In Fig.~2, this transition is seen as an
anticrossing between the ground hole state and the first excited state, which also accounts
for the anomalous quantum confined Stark effect.

Although the anomalous quantum confined Stark effect invalidates Eq.~(1) over the whole range
of electric field, it is possible to write two separate equations that describe the dependence
of the transition energy on electric field for the stacked structure, except for the electric
fields at which anticrossings occur.

\begin{eqnarray}
&E^+(F) = E(0) + p^+ F + \beta^+ F^2,& \nonumber \\
&E^-(F) = E^\prime + p^- F + \beta^- F^2.&
\end{eqnarray}
where $E^\prime$ is a fitting parameter, $E^+(F)$ is for $F>-21.4$~kV/cm, and $E^-(F)$ is
for $F<-21.8$~kV/cm.

For the right branch, $p^+/|e|=-3.65$~nm is the built-in dipole moment of the stacked
structure at zero field. From the middle panel of Fig.~4(a), the ground electron and hole
states are seen to be separated by a distance larger than the coupling region ($1.8$~nm), and
much larger than the electron-hole separation in the single dot, which results in a dipole
moment of more than one order of magnitude larger than that in the single dot. For the left
branch, $p^-\approx -p^+$~nm has a positive sign because of the inverted electron-hole
alignment \cite{swd} in the stacked structure at negative electric fields.

The quantum confined Stark effect coefficient is defined by $\beta=\beta_e-\beta_h$ where
$\beta_e$ ($\beta_h$) can be related to oscillator strength of optical intraband transitions
in conduction (valence) bands \cite{bar,jpl}. For instance, $\beta_e$ is given by

\begin{equation}
\beta_e = -\frac{\hbar^2}{2m_0}\sum_{n>1} f_{1\rightarrow n}/(E_n-E_1)^2 ,
\end{equation}
where $m_0$ is the bare electron mass, $f_{1\rightarrow n}$ is the oscillator strength for the
intraband transition from the ground state to the n-th state, with polarization along $z$
direction. $\beta_h$ has a similar expression. Single SAD structures have been shown to have
very weak $z$-polarized intraband transitions \cite{apl}, therefore both $\beta_e$ and
$\beta_h$ are small. In the stacked structure, the strengths of the $z$-polarized intraband
transitions in the valence bands are similar to those in the single dot. {\it i.e.}, $\beta_h$
is small, which could be observed in Fig.~2 where the hole state energies show a quasi-linear
dependence on the electric field. However, in the conduction bands of the stacked structure,
there are several strong intraband transitions, especially the $1s^+\rightarrow 1s^-$
transition \cite{apl}. This results in a much larger $\beta_e$ than in single dots, and
explains the magnitude of the quantum confined Stark effect coefficients in both branches
($\beta^\pm/|e|^2\approx -80\mbox{~nm}^2/eV$) which are six times larger than in single dots,
and are responsible for the `bowing' of the Stark shift in Fig.~1.

In conclusion, we have shown that vertically stacked InAs/GaAs self-assembled quantum dots
exhibits, in addition to a much larger Stark shift than single dots, a strong non-parabolic
dependence of interband transition energy on the electric field, not been encountered in other
types of quantum structures. We have demonstrated that the 3D distribution of the biaxial
strain field is mainly responsible for this anomalous quantum confined Stark effect.

\acknowledgments
This work is supported by Army Research Office, grant \#DAAD 10-99-10129 and National Computational
Science Alliance, grant \#ECS000002N.


\begin{references}
\bibitem{ft1} In a one-dimensional triangular potential created by an electric field $F$, the
ground state is an Airy function with energy $E \propto F^{4/3}$. Here, we found $E \propto
F$, which is due to the 3D nature of the potential, as well as to the fact that its profile is
not entirely linear.
\end{references}
\end{document}